\documentclass[prl,aps,twocolumn,amsmath,amssymb,showpacs]{revtex4}

\usepackage{graphicx}
\usepackage{epsfig}
\usepackage[english]{babel}
\usepackage{natbib}
\usepackage{amsfonts}
\usepackage{amssymb}
\usepackage{graphicx}
\usepackage{dcolumn}
\usepackage{amsmath}
\usepackage[ansinew]{inputenc}
\usepackage{psfig}
\usepackage{amssymb}

\begin{document}
\date{\today}
 \title
{Chaotic dynamics of superconductor vortices in the plastic phase}
\author
{E. Olive, J.C. Soret}
\affiliation{LEMA, UMR 6157, Universit\'e Fran\c{c}ois Rabelais - CNRS - CEA, Parc de Grandmont 37200 Tours, France}


\begin{abstract}
We present numerical simulation results of driven vortex lattices in presence of random disorder at zero temperature. 
We show that the plastic dynamics is readily understood in the framework of chaos theory. Intermittency "routes to chaos" have been clearly identified, and positive Lyapunov exponents and broad-band noise, both characteristic of chaos, are found to coincide with the differential resistance peak. Furthermore, the fractal dimension of the strange attractor reveals that the chaotic dynamics of vortices is low-dimensional. 
 \end{abstract}

\maketitle

When flowing over a random medium, vortices in type II superconductors display a great variety of dynamical regimes, from the depinning threshold up to the high driving phase. 
Most of the  $V-I$ experiments \cite{Higgins,manip} and numerical simulations \cite{Gronbech, FaleskiRyuOlson,Spencer,Kolton,simul} reveal an intricate interplay between the 'peak effect' (PE), {\it i.e.} the increase of the depinning threshold current below the upper critical field $H_{c2}$, the peak of the differential resistance $dV/dI$, voltage noise and plastic flow of vortices. Below the PE an ordered phase is expected and the unusual excess noise measurements are understood within an edge contamination process where a metastable disordered vortex phase generated at the edges is annealed into an ordered phase in the bulk \cite{Paltiel}. On the contrary, in the PE region a disordered phase is expected and plasticity effects such as tearing are expected at the depinning threshold. These features have recently been studied in the mean field approach \cite{Marchetti}. However many open questions about the complex plastic flow remain, and in particular its dynamical properties.\\
In this Letter, we propose to examine the plastic phase through the chaos theory of deterministic dissipative dynamical systems. Charge density waves and Josephson junctions arrays have already been analysed through chaos theory \cite{LevyMarconi}, but such study is completely new for vortex lattices. We performed numerical simulations that clearly demonstrate the chaotic behavior of vortices in the plastic phase. While increasing the driving force, instabilities are developed by the non-linearities of the system and periodic regimes are destabilized giving rise to chaotic regimes. Such destabilizations have been clearly identified in our system to be the intermittency "route to chaos". Furthermore, the broad-band voltage noise, the positive Lyapunov exponents and the fractal dimension of the strange attractor are used to characterize the chaotic phase which is shown to coincide with the peak of $dV/dI$. A crucial result of our study shows that the chaotic dynamics in the plastic phase is low-dimensional.
Therefore within the framework of chaotic dynamical systems, our results open new perspectives in the understanding of vortex dynamics that are discussed in the conclusion.

We consider $N_v$ Abrikosov vortices driven over a random pinning background in the $(x,y)$ plane.
At $T=0$ the overdamped equation of motion of a vortex $i$ in position $\bold r_i$ reads 
   \begin{eqnarray}
\eta {{d{\bf r_i}}\over{dt}}=-{\sum_{j \neq i}}\nabla_i U^{vv}(r_{ij})-{\sum_{p}}\nabla_i U^{vp}(r_{ip})+{\bf F}^L
\label{eq1}  
 \end{eqnarray}
where $r_{ij}$ is the distance between vortices $i$ and $j$, $r_{ip}$
 is the distance between the vortex $i$ and the pinning site located at ${\bf r_p}$, and $\nabla_i$ is the 2D gradient operator acting in the 
$(x,y)$ plane. The viscosity coefficient is $\eta$, ${\bf F}^L=F^L{\bf \hat x}$ is the Lorentz driving force due to an applied current.
The vortex-vortex pairwise repulsive interaction is given by a modified Bessel function $U^{vv}(r_{ij})=2\epsilon_0A_v K_0(r_{ij}/\lambda_L)$, and the attractive pinning potential is given by $U^{vp}( r_{ip})=-\alpha_p e^{-(r_{ip}/R_p)^2}$. In these expressions $A_v$ and $\alpha_p$ are tunable parameters, $\lambda_L$ is the magnetic penetration depth, and $\epsilon_0=(\phi_0/4\pi \lambda_L)^2$ is an energy per unit length. 
We consider periodic boundary conditions of $(L_x, L_y)$ sizes in the $(x,y)$ plane. All details about our 
method for computing the Bessel potential with periodic conditions can be found in \cite{Olive}. Molecular Dynamics simulation is used for $N_v=30$ 
vortices in a rectangular basic cell $(L_x,L_y)=(5, 6\sqrt 3/2)\lambda_L$. 
The number of pinning centers is set to $N_p=30$. We consider the London limit $\kappa =\lambda_L /\xi =90$, where $\xi$ is the superconducting coherence length \cite{note1}. The average vortex distance $a_0$ is set to $a_0=\lambda_L$, and  $R_p=0.22\ \lambda_L$, $\eta=1$, $A_v=2.83\times 10^{-3}\lambda_L$. We present results for two different pinning strengths
corresponding to a maximum pinning force of $F_{max}^{vp}\sim 0.2F_0$ and $F_{max}^{vp}\sim 1.4F_0$,  
where $F_0=2\epsilon _0A_v/\lambda_L$ is a force defined by the Bessel interaction.
In the weak pinning case, the driving force applied along a principal vortex lattice direction $x$ is varied from $0$ up to $F^L\sim 3F_0\sim 100F_c^L$, where $F_c^L$ is the critical Lorentz force along $x$. In the strong pinning case, the driving force is varied from $0$ up to $F^L\sim 3F_0\sim 20F_c^L$.
The choice of the double precision Runge-Kutta algorithm time iteration step $\delta t$ is dictated by the dominant force, and for example in the high driving phase we take $\delta t=10^{-3}t_0$, where $t_0=\eta\lambda_L/F^L_{max}$.

The "experimental" procedure is the folllowing: we start by randomly throwing in the $(x,y)$ plane $N_v$ vortices and $N_p$ gaussian pins, 
and relaxation with zero Lorentz force yields a vortex structure with dislocations. The Lorentz force is then slowly increased up to far in the high driving phase. 
The successive regimes we observe in the weak pinning case $F_{max}^{vp}\sim 0.2F_0$
are the following. Phase I: pinned regime where all vortices have zero velocity. Phase II: plastic channels flowing through pinned regions and where the motion is either periodic or quasiperiodic as seen in \cite{Gronbech}. Phase III: 
plastic flow with almost no stationnary vortices as seen in \cite{Gronbech, Fangohr,Spencer}. In the following this motion shall be shown to be chaotic. Phase IV: fully elastic flow with no dislocations and where the motion occurs through rough static channels \cite{Giamarchi}.\\ 
We shall now examine in details the plastic dynamics of vortices in the framework of classical chaos theory. The first central point of the letter shows that the transition from phase II to phase III is one of the three well-known "routes to chaos".
In this very short applied force range, the typical longitudinal velocity of the vortex center of mass $V_x^{cm}$ that we measure in time is shown in  Fig.\ 1a.
\begin{figure}
\includegraphics[width=0.86\linewidth]{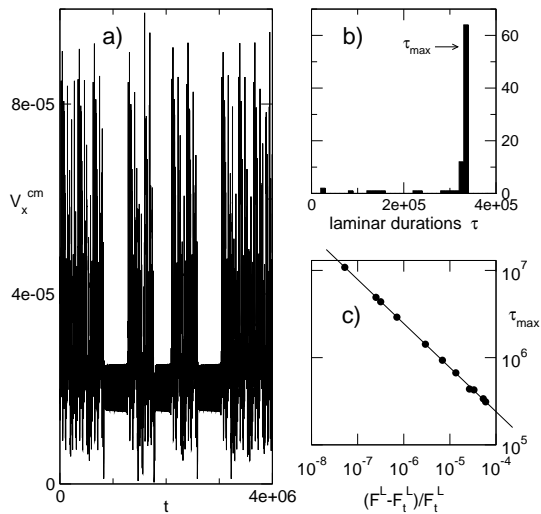}
\caption{ \label{fig1} Properties of the transition region from phase II to phase III in the weak pinning case. It clearly displays the type I intermittency route to chaos characteristics. a) part of the time evolution of the longitudinal velocity $V_x^{cm}(t)$ obtained for $F^L=1.5221\times 10^{-4}$. One sees laminar ({\it i.e.} periodic) phases interupted by chaotic bursts of large velocity fluctuations. b) distribution of the laminar phase durations of $V_x^{cm}(t)$ measured for $F^L=1.5221\times 10^{-4}$. 
c) evolution of $\tau_{max}$ when varying the applied force $F^L$ from the intermittency threshold force $F_t^L$. A clear power law is observed as shown with the line of slope $-1/2$.}
\end{figure}
 It shows time intervals where the motion is periodic (the same as used to exist in phase II below the transition). The difference is now that such periodic regime becomes unstable and gives way to a chaotic burst displaying large velocity fluctuations. Then the system goes back to the periodic regime which is still unstable, giving way to another chaotic burst, and so on. The chaotic bursts correspond to apparently disordered trajectories of the moving vortices. However, from time to time, moving vortices are able to synchronize temporarily their motion into 
periodic motion (laminar phases). In the framework of the dissipative chaos theory, such intermittent regimes are known to be one possible way to drive the system from periodicity to chaos. The intermittency "route to chaos" has several characteristics and may be mainly classified in three types (I, II and III) depending on the unit circle crossing value of the Floquet multipliers \cite{Pommeau,Berge}. To determine the type of intermittency we observe in our system, we first measure for a given value of the applied force the distribution of the laminar ({\it i.e.} periodic) phase durations. Fig.\ 1b shows such distribution obtained for $V_x^{cm}(t)$ displayed in Fig.\ 1a. This distribution of laminar phase durations shows a maximum at an upper bound $\tau_{max}$ and a decrease for low durations which can be much smaller than $\tau_{max}$. Furthermore, if we now increase very slowly the applied force in order to remain in an intermittent regime, the value of $\tau_{max}$ decreases as shown in Fig.\ 1c. A very nice power law $\tau_{max}\sim (F^L-F_t^L)^{-1/2}$ on almost four decades is measured close to the intermittency threshold $F_t^L$, {\i.e.} the force above which periodic regimes become unstable. The particular shape of the distribution of the laminar phase durations and the exponent $-1/2$ are characteristics of type I intermittency route to chaos related to a saddle-node bifurcation at $F_t^L$ \cite{Pommeau,Berge}. Note that the type of intermittency may change for different pinning strengths (for stronger pinning parameters $F_{max}^{vp}\sim 1.4F_0$, we observed for example a type II intermittency route to chaos characterized in particular by a different shape of the distribution of the laminar phase durations, and related to a subcritical Hopf bifurcation). Further increasing the applied force will give intermittent regimes with shorter laminar phase durations until they completly disappear therefore giving way only to large chaotic fluctuations. Then chaos expands in phase III.

\begin{figure}
\includegraphics[width=0.86\linewidth]{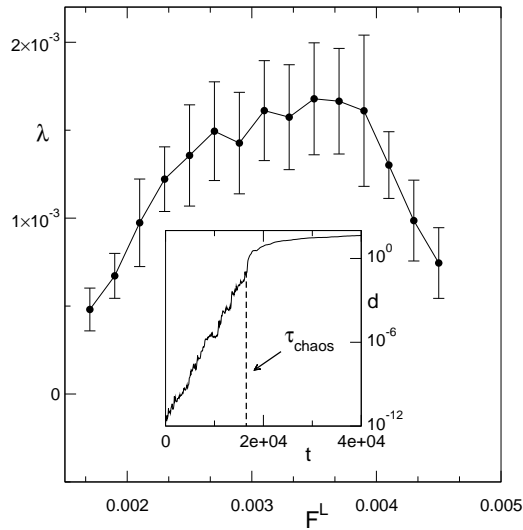}
\caption{ \label{fig2} Evolution of the maximal Lyapunov exponent $\lambda$ with the Lorentz force in phase III. Each point is the average of 20 couples of initial conditions and the error bars are the standard deviation. The inset displays the time evolution of the distance $d(t)$ between two initial neighbouring trajectories in the phase space for $F^L=0.0029$ in the strong pinning case. The exponential divergence (positive slope $\lambda$) characteristic of chaos is obvious for $t<\tau_{chaos}$. For time scales larger than $\tau_{chaos}$ diffusive motion is observed as already shown in \cite{Kolton}. }
\end{figure}

\begin{figure}
\includegraphics[width=0.86\linewidth]{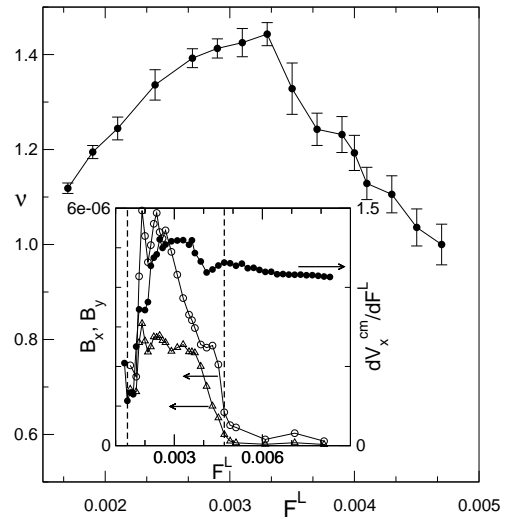}
\caption{ \label{fig3} Evolution of the strange attractor fractal dimension (computed with the correlation dimension $\nu$) with the Lorentz force in phase III. The inset displays the differential resistance curve (filled circles) and the low frequency longitudinal $B_x$ (open circles) and transverse $B_y$ (triangles) noises. Dotted lines separate phases II, III and IV. 
}
\end{figure}

In the second central point of our letter, we examine in details the chaotic phase itself. Usual tools of chaos theory are successfully used to characterize the chaotic dynamics of vortices, and the link with the commonly observed differential resistance peak \cite {Higgins,Kolton,FaleskiRyuOlson} in vortex dynamics is established. This section characterizes the chaotic attractor of the vortices in the plastic phase III of the strong pinning case $F_{max}^{vp}\sim 1.4F_0$ (where phases I, II and III are equivalent to those described above for the weak pinning case). 
We first compute the Lyapunov exponents. Positive Lyapunov exponents are a signature of chaotic dynamics since they illustrate the "sensitive dependence on initial conditions" (SDIC) which is a property of chaotic attractors only.
To compute the maximal Lyapunov exponent of our system we consider two very close initial conditions and observe how the distance $d(t)$ in the phase space between the two corresponding trajectories evolve in time on the attractor. Since we integrate $N_v$ first order differential equations of motion (Eq.\ \ref{eq1}), the phase space is defined by the $2N_v$ vortex coordinates 
and the distance $d$ is defined by
$d^2(t)={\sum_{i=1}^{N_v}}\left[\left(X_i(t)-\tilde X_i(t)\right)^2+\left(Y_i(t)-\tilde Y_i(t)\right)^2\right]$ where $X_i(t)=x_i(t)-x_{cm}(t)$, $Y_i(t)=y_i(t)-y_{cm}(t)$, $\tilde X_i(t)=\tilde x_i(t)-\tilde x_{cm}(t)$, $\tilde Y_i(t)=\tilde y_i(t)-\tilde y_{cm}(t)$. In these expressions, $(x_i,\ y_i)$ and $(\tilde x_i,\ \tilde y_i)$ are the vortex $i$ coordinates, 
and $(x_{cm},\ y_{cm})$ and $(\tilde x_{cm},\ \tilde y_{cm})$ are the respective coordinates of the center of mass. The tilde notation ($\tilde x,\tilde y$) refers to the second trajectory generated by the neighbouring initial condition. 
The inset of Fig.\ \ref{fig2} displays an example of the time evolution of $d$ we typically find in phase III. 
It clearly shows an exponential divergence $d\sim \exp(\lambda t)$ of the two trajectories for time scales up to $\tau_{chaos}\sim 1.6\times10^4\sim 50t_0$. The slope therefore defines a positive maximal Lyapunov exponent $\lambda$ characteristic of chaotic dynamics. Fig.\ \ref{fig2} displays the evolution of $\lambda$ that we observe in phase III. The maximal value of $\lambda$ therefore expresses that the fastest divergence of two chaotic trajectories occurs in the midrange of phase III.  
Finally, note that for time scales larger than $\tau_{chaos}$ we find diffusive and superdiffusive motions (not shown) in the transverse and longitudinal directions as already reported in \cite{Kolton}. 
For a given applied force we now compute the power spectrum $S_{\alpha}(f)$ of $V_{\alpha}^{cm}(t)$ , {\it i.e.} $S_{\alpha}(f)={{1}\over{t_2-t_1}} \mid \int_{t_1}^{t_2}dtV_{\alpha}^{cm}(t)\ exp(i2\pi ft )\mid ^2$ where $\alpha=x$ or $y$, and $t_2-t_1 \gg \tau_{chaos}$. We define the low frequency noise $B_{\alpha}$ by averaging $S_{\alpha}(f)$ over the low frequency range \cite{note2}. $B_x$ and $B_y$ are in some way a measure of the degree of chaos in the system since chaotic dynamics generates broad-band noise at low frequencies \cite{Berge}. Concomitantly with the differential resistance $dV_x^{cm}/dF^L$we plot in the inset of Fig.\ \ref{fig3} the longitudinal $B_x$ and transverse $B_y$ low frequency noises. 
The rapid increasing of $B_x$ and $B_y$ confirm the (rapid) setting of chaos in the vortex lattice and their maximal value coincide with the well-known peak of the differential resistance. As shown in Fig.\ \ref{fig2}, it also corresponds to the maximal value of $\lambda$. Chaos is therefore fully developed at the differential resistance peak.\\ 
The positive Lyapunov exponents characrerize the absence of temporal correlation in the chaotic regime due to SDIC. We shall now characterize the spatial correlations within the chaotic regime by computing the dimension of the chaotic attractor. Such an attractor is known to be fractal with a non-integer dimension in the phase space. To characterize the fractal nature of this so-called strange attractor, we evaluate the correlation sum defined by $C(\rho)=lim_{m\rightarrow \infty}1/m^2{\sum_{k,l=1}^m}H(\rho-\rho_{kl})$ which measures the number of couples of points $(k,l)$ on the chaotic attractor whose distance $\rho_{kl}$ is less than $\rho$. $H(z)$ is the Heaviside function. For a limited range of $\rho$ it is found that $C(\rho)\sim \rho^{\nu}$ where the exponent $\nu$ is called the correlation dimension and is a simple measure of the fractal dimension of the attractor \cite{Grassberger}. As shown in Fig.\ \ref{fig3}, the fractal dimension of the vortex strange attractor has the same shape as the Lyapunov exponent curve (Fig.\ \ref{fig2}) and its maximum also coincides with the differential resistance peak (inset of Fig.\ \ref{fig3}). It therefore confirms our previous finding that chaos is fully developed at the peak of the differential resistance curve. Above the peak, chaos still exists but its intensity decreases as shown by the deacreasing of $\nu, \lambda, B_x$ and $B_y$. It corresponds to the onset of transverse order of the smectic phase \cite{Balents} which we find to occur precisely at the peak. 
Furthermore, we find the very important result that the strange attractor of the driving vortices has a low (fractal) dimension, $1<\nu<2$ (Fig.\ \ref{fig3}), which shows that the chaotic dynamics of a large number of vortices shrinks on a low-dimensional surface in the phase space. The crucial consequence of a strange attractor of dimension less than two is that the chaotic dynamics of vortices in the plastic phase may be described with only three dynamical variables. This is an important result for further theoritical studies because simple analytical models with three dynamical variables should be sufficient to describe the complex plastic phase.
Finally, we find that the end of chaos coincides with the dynamical freezing transition \cite{Kolton} where the transverse velocity $V_y^{cm}$ drops to zero. We therefore clearly showed that the bottom of the differential resistance peak marks the onset of chaos, while the end of the peak and the dynamical freezing transition appear as the end of chaos.

In conclusion, we obtained conclusive results about the chaotic dynamics of vortices in the plastic phase. The route to chaos has been identified in details: type I (II) intermittency in the weak (strong) pinning case is found. Chaos characterized by positive Lyapunov exponents and broad band noise is found to coincide with the differential resistance peak. Furthermore, the fractal dimension of the strange attractor shows that the chaotic dynamics of vortices is low-dimensional. 
Therefore our results open new perspectives in the theoretical understanding of the plastic flow phase which is much less developed than the fast moving vortex phases \cite{Giamarchi,Balents} and than the plastic depinning transition \cite{Marchetti}. In particular we show the important result that the chaotic dynamics in the plastic phase may be understood with only three dynamical variables. Hence, our results combined with the usual tools of the chaos theory ({\it e.g.} bifurcation theory, embedding dimensions, Poincaré sections, strange attractors, time series analysis) should help for further theoretical, numerical and experimental investigations of the open issues related to the formation of fractal objects in the complex space phase of driven plastic vortices.
Finally, our results let foresee new possibilities of controlling vortex motion for device applications
using the concept of {\it controlling chaos} developed these last years (see {\it e.g.} \cite{Abarbanel}). 
The goal of such control procedure is to lock the chaotic system into a stable periodic orbit 
which either used to be unstable in the uncontrolled system (feedback scheme by weakly changing parameters \cite{Ott}), or which is newly created (non-feedback scheme by weakly forcing the system, see {\it e.g.} \cite{Braiman}).  
We therefore suggest that the concept of controlling chaos might be used to design device applications to rectify plastic (chaotic) vortex motion.

\begin{acknowledgments}
We are grateful to the LMPT-CNRS UMR 6083 - Tours (France) for our extensive use of their computers. We aknowledge discussions with T. Giamarchi, P. Le Doussal, A. Mouchet, H. Giacomini and Y. Lansac.
\end{acknowledgments}


\end{document}